\documentclass[preprint,aip,rsi,amsmath,amssymb]{revtex4-1}
\usepackage{graphicx}

\begin{document}
\title{Fully integrated InGaAs/InP single-photon detector module with gigahertz sine wave gating}
\author{Xiao-Lei Liang}
\affiliation{Hefei National Laboratory for Physical Sciences at Microscale and Department of Modern
Physics, University of Science and Technology of China, Hefei, Anhui 230026, China}

\author{Jian-Hong Liu}
\affiliation{Anhui Quantum Communication Technology Co., Ltd., Hefei, Anhui 230088, China}

\author{Quan Wang}
\affiliation{Anhui Quantum Communication Technology Co., Ltd., Hefei, Anhui 230088, China}

\author{De-Bing Du}
\affiliation{Anhui Quantum Communication Technology Co., Ltd., Hefei, Anhui 230088, China}

\author{Jian Ma}
\affiliation{Hefei National Laboratory for Physical Sciences at Microscale and Department of Modern
Physics, University of Science and Technology of China, Hefei, Anhui 230026, China}

\author{Ge Jin}
\affiliation{Hefei National Laboratory for Physical Sciences at Microscale and Department of Modern
Physics, University of Science and Technology of China, Hefei, Anhui 230026, China}

\author{Zeng-Bing Chen}
\affiliation{Hefei National Laboratory for Physical Sciences at Microscale and Department of Modern
Physics, University of Science and Technology of China, Hefei, Anhui 230026, China}

\author{Jun Zhang}
\email{zhangjun@ustc.edu.cn}
\affiliation{Hefei National Laboratory for Physical Sciences at Microscale and Department of Modern
Physics, University of Science and Technology of China, Hefei, Anhui 230026, China}

\author{Jian-Wei Pan}
\affiliation{Hefei National Laboratory for Physical Sciences at Microscale and Department of Modern
Physics, University of Science and Technology of China, Hefei, Anhui 230026, China}
\date{\today}

\begin{abstract}
InGaAs/InP single-photon avalanche diodes (SPADs) working in the regime of GHz clock rates are
crucial components for the high-speed quantum key distribution (QKD). We have
developed for the first time a compact, stable and user-friendly tabletop InGaAs/InP single-photon detector system operating at a 1.25 GHz gate rate
that fully integrates functions for controlling and optimizing SPAD performance.
We characterize the key parameters of the detector system and test the long-term
stability of the system for continuous operation of 75 hours. The detector system
can substantially enhance QKD performance and our present work paves the way for practical high-speed QKD applications.
\end{abstract}
\maketitle
\section{Introduction}
For applications requiring near-infrared single-photon detection such as quantum key distribution (QKD) \cite{GRTZ02},
optical time domain reflectometry \cite{OTDR} and eye-safe laser ranging \cite{LR}, InGaAs/InP single-photon
avalanche diodes (SPADs) working in the Geiger mode are widely used due to their practicality \cite{INGAAS07}.
Numerous works on quenching techniques and afterpulsing reduction for single-photon detection have been performed over the
past two decades. Afterpulsing refers to undesired subsequent avalanches that are due to the depopulation of the trapped
charge carriers created during previous avalanches.
Suppressing the afterpulsing effect is a key step to improve SPAD performance.
The afterpulse probability can be roughly modeled as \cite{ZTGGZ09}
\begin{equation}
\label{pap}
P_{ap}\propto (C_{d}+C_{par})\int_{0}^{\Delta t}V_{ex}(t)dt \times e^{-T_{d}/\tau},
\end{equation}
where $C_{d}$ is the diode capacitance, $C_{par}$ is the parasitic capacitance, ${\Delta t}$ is the avalanche
duration time, $V_{ex}$ is the excess bias, $T_{d}$ is the deadtime, and $\tau$ is the lifetime of detrapping
carriers. According to Eq. \ref{pap}, there are several approaches to reduce
afterpulsing. Increasing the operating temperature ($T$) in order to decrease $\tau$, or increasing $T_{d}$, can both reduce the release rate of the
trapped carriers. However, higher temperatures deteriorate the dark count performance while longer deadtime settings limit
the maximum count rate. Minimizing $C_{par}$ can reduce afterpulsing by decreasing the avalanche charge quantity.
This has been demonstrated, for instance, in experiments using integrated quenching electronics \cite{ZTGGZ09} or the hybrid packaging technique \cite{PQAR08}. Nevertheless, these sophisticated techniques are challenging.
Significantly reducing ${\Delta t}$ can also effectively limit the avalanche charge quantity,
which is the original idea of the recently emerging technique called rapid gating \cite{NSI06,NAI09,GAP09,GAP10,GAP10-2,NEC11,Toshiba07,Toshiba10,Italy12}.

Once the afterpulsing effect is significantly suppressed, InGaAs/InP SPADs can operate in the high-speed mode.
For applications like QKD, high repetition frequency is critical for key generation. Such high-speed
QKD experiments using rapid gating detectors have recently been demonstrated \cite{QKD1,QKD2,Nihon07,Nihon10}.
In rapid gating schemes, due to an ultrashort gating duration, avalanche charge quantity is extremely limited.
As a result, the afterpulsing effect is sufficiently reduced, and the gating frequency can easily reach the GHz level.
Low charge flow however, creates a weak avalanche signal to be
detected. There are currently two approaches to extract weak avalanches from huge capacitive response signals:
sine wave gating and filtering \cite{NSI06,NAI09,GAP09,GAP10,GAP10-2,NEC11},
and self-differencing \cite{Toshiba07,Toshiba10}.

\section{Detector system design}

\begin{figure}[t]
\centering
\includegraphics[width=7.6 cm]{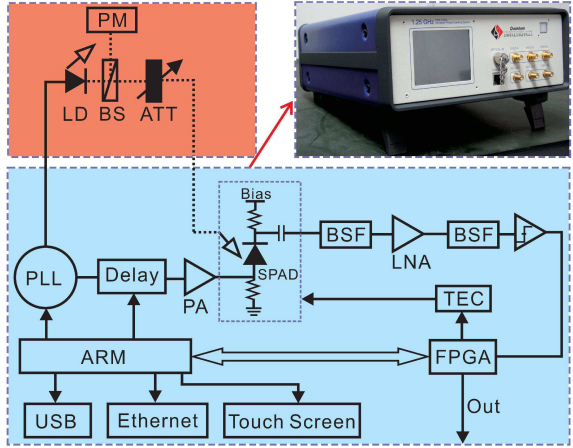}\\
\caption{Experimental setup for single-photon characterization (top left), detector system photo (top right), and detector system diagram (bottom).
See text for explanation.}
\label{fig1}
\end{figure}

In order to be effectively and reliably applied in practical QKD applications, a compact and stable rapid gating detector is indispensable.
We have developed for the first time a fully integrated tabletop InGaAs/InP single-photon detector working at
a gating frequency of 1.25 GHz (see Fig. \ref{fig1}). The practical features of the detector system include:
multiple independently programmable
high precision clock outputs, internal and external clock references, gate amplitude control,
low-noise fine tuning bias, automatic delay adjustment for synchronized photon detection,
switchable avalanche signal monitor, adjustable SPAD temperature,
tunable discrimination threshold, programmable deadtime, and user-friendly interface.
All system components are assembled into a module 25 cm$\times$ 10 cm$\times$ 33 cm in size as shown in Fig. \ref{fig1}.

\begin{figure}[t]
\centering
\includegraphics[width=7.6 cm]{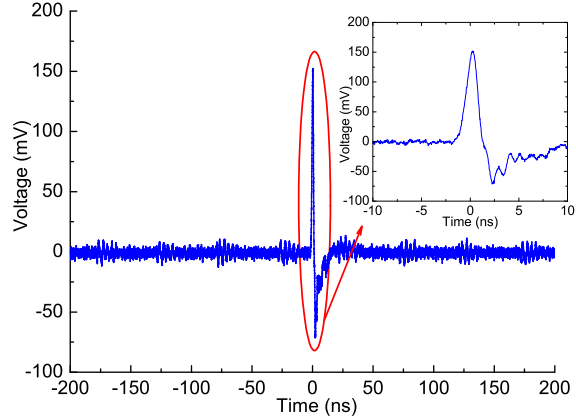}\\
\caption{Oscilloscope trace of the avalanche signal. Inset: expanded view of the main peak.}
\label{fig2}
\end{figure}

The rapid gating scheme of the detector system is based on sine wave gating and filtering \cite{NSI06,NAI09,GAP09,GAP10,GAP10-2,NEC11}.
A phase-locked loop (PLL) generates initial sine waves
at a frequency of $f_g$=1.25 GHz. Using a power amplifier (PA) with a gain of 40 dB, the sine waves have a peak-peak
amplitude ($V_{pp}$) of 10 V. $V_{pp}$ may be further adjusted
by a digital attenuator. The SPAD response signal initially passes through microwave band-stop filters (BSFs) at
a center frequency of 1.25 GHz. The attenuated signal is then amplified by a low-noise
amplifier (LNA) and subsequently filtered by the second set of BSFs. The capacitive response signal is then
suppressed to a negligible level where the avalanche signal can be discriminated.
An electronic switch also allows the avalanche signal to be monitored on an oscilloscope.

Fig. \ref{fig2} shows a typical avalanche trace captured
on a 12 GHz oscilloscope (Agilent infiniium). The avalanche amplitude is greater than 100 mV with the peak-peak electronic noise
below 20 mV. This high signal-to-noise ratio is attributed to careful consideration
of signal integrity in our circuit board design, good electrical shielding, and ultra-low noise bias
with a root mean square ripple less than 800 $\mu$V. The undershoot of the avalanche
trace is due to the capacitive effect, since the output from SPAD is alternating current coupled with a capacitor.

In order to achieve the objectives of easy-to-use and stable long-term operation,
auxiliary hardware and software designs are used to control the detector system.
The control center is an advanced RISC machine
(ARM) embedded system. The user interface is an on-board touch screen that communicates
with an embedded system program developed in Nokia's Qt framework.
Alternatively, a computer can control the detector system either via
universal serial bus (USB), or remotely via ethernet. A field-programmable gate array (FPGA)
and a thermoelectric cooling (TEC) driver are used to maintain SPAD temperature.
FPGA can also perform counting, and apply logic deadtime to the output of the discriminator.
Logic deadtime refers to the ``count-off time'' as discussed in Ref. \cite{GAP09,GAP10}.
Unlike real deadtime, during which the SPAD is not operating in the Geiger mode, logic deadtime
only prevents counting for a period after a primary detection \cite{GAP09,GAP10}.

\section{Characterization and discussion}

\begin{figure}[t]
\centering
\includegraphics[width=7.6 cm]{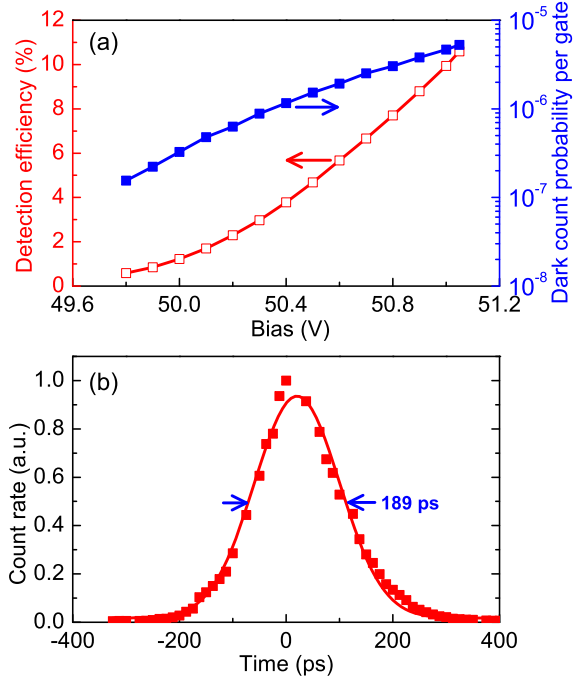}\\
\caption{(a) Detection efficiency ($\eta$) and dark count probability per gate ($P_{dc}$) as a function of bias.
(b) Effective gating width measurement by scanning the relative delay between optical pulses and sine gates at $\eta$=10 \%.}
\label{fig3}
\end{figure}

For SPAD characterization, an optical test bench is constructed as shown in Fig. \ref{fig1}.
A laser diode (LD, picoquant PDL 800-D) is triggered by the PLL synchronized output at a
frequency of $f_l$=$f_g$/100. The ultrashort optical pulses are divided by an asymmetric beam splitter (BS).
The optical power in one of the BS output ports is monitored by a calibration power meter (PM, IQS-1500).
A LabVIEW program reads the measurement results of the PM and regulates a precise variable attenuator
(ATT, IQS-3150) in real time in order to cancel the intensity fluctuations of the LD.
The intensity of the resulting optical pulses reach the single-photon level.

Fig. \ref{fig3}(a) shows the efficiency-noise performance of the SPAD under the operating conditions of
$T$=-50 $^{\circ}$C and a mean photon number per optical pulse of $\mu$=1. As the bias rises the system detection
efficiency ($\eta$) increases linearly while the dark count probability per gate ($P_{dc}$) increases exponentially.
When $\eta$=10 \%, $P_{dc}$ is $4.66\times 10^{-6}$, corresponding to 5.83 kHz dark count rate. The effective gating width at $\eta$=10 \%
is measured by scanning the relative delay between optical pulses and sine gates. The results are plotted and
fitted in Fig. \ref{fig3}(b). The full width at half maximum of the gating width is 189 ps, and the resulting
normalized dark count probability per ns at $\eta$=10 \% is $2.47\times 10^{-5}$. This is consistent with
the results that are obtained with the same InGaAs/InP SPAD using a conventional low-frequency gating technique (10 kHz gating frequency
and 10 ns gating width) under the same conditions of $T$ and $\eta$.
This similarity suggests that dark count performance of SPAD is
independent of the quenching techniques.

\begin{figure}[t]
\centering
\includegraphics[width=7.6 cm]{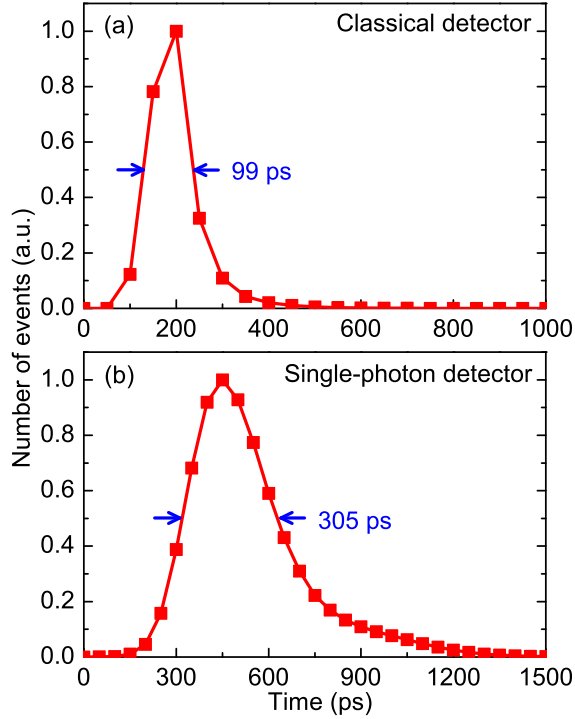}\\
\caption{(a) The overall jitter of LD, PLL, and the TDC electronics measured with a 12.5 GHz InGaAs PIN detector. The jitter of the classical detector is negligible. (b) The overall jitter measured with the rapid gating InGaAs single-photon detector at $\eta$=10 \%.}
\label{fig4}
\end{figure}

Timing resolution is a key parameter of SPAD for high-speed applications.
The intrinsic jitter of the system is measured with a 12.5 GHz InGaAs PIN photodetector (ET-3500F)
using a multistop time to digital converter (TDC, Agilent Acqiris)
of 50 ps timing resolution. The overall jitter including the contributions of LD, PLL,
and TDC is 99 ps as shown in Fig. \ref{fig4}(a). The jitter of the whole detector system with the rapid gating SPAD at $\eta$=10 \%
is measured to be 305 ps (see Fig. \ref{fig4}(b)). Therefore, the timing resolution of the rapid gating SPAD
itself is $\sqrt {305^{2}-99^{2}}=288$ ps. There are two reasons for this moderate timing resolution. First,
this resolution includes contributions from the discrimination circuit and the FPGA, which are difficult to quantify separately.
Second, in the rapid gating scheme, although the cascaded BSFs minimize capacitive response signals, they can also distort the avalanche signals.
Such distortions add timing jitter \cite{Zeng11}.

\begin{figure}[t]
\centering
\includegraphics[width=7.6 cm]{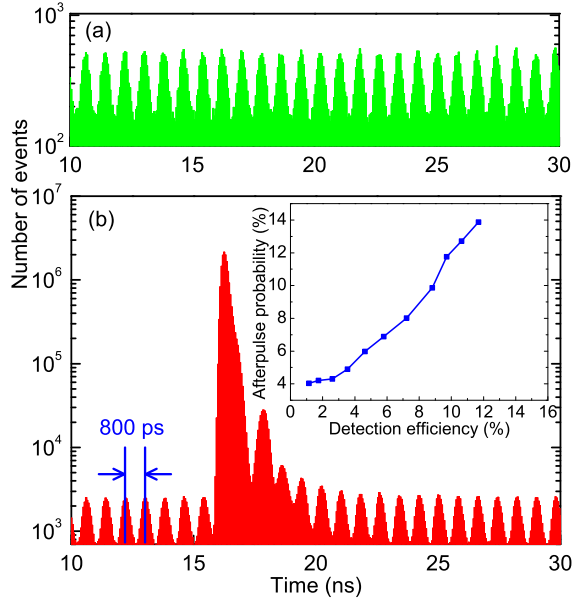}\\
\caption{The histogram of detection events under the conditions of
$\eta$=10 \%, $\mu$=0.1, $T$=-50 $^{\circ}$C and a 10-minute integration time
without (a) and with (b) optical illumination. Inset: $P_{ap}$ versus $\eta$ using the same bias
settings as in Fig. \ref{fig3}(a).}
\label{fig5}
\end{figure}

For afterpulsing characterization, although the standard double-gate method \cite{ZTGGZ09} is the most precise approach,
it cannot be applied to rapid gating. In our experiment we use the high-speed TDC
system to directly tag the timing of all the avalanche detections and then statistically analyze the events offline.
At each bias setting we acquire the distribution of detection events both with and without optical illumination.
The common measurement conditions include $f_l$=12.5 MHz, $\mu$=0.1, $T$=-50 $^{\circ}$C, a 10-minute integration time,
and without applying logic deadtime \cite{GAP09,GAP10}. Histograms for $\eta$=10 \% are shown in Fig. \ref{fig5}.
Fig. \ref{fig5}(a) shows an equal distribution of dark counts in which peaks are separated by 800 ps.
Fig. \ref{fig5}(b) shows the corresponding histogram with the illumination of the pulsed laser.
In this case, the main peak is due to photon detections and the subsequent decay shows the typical afterpulsing effect.
For practical applications appropriate coincidence technique may be necessary to clearly separate the main peak and
the first side peak.

In the experiment, the total afterpulse probability $P_{ap}$ is calculated using
\begin{equation}
\label{pa}
P_{ap}=\frac{C_{tol}-C_{dc}-C_{ph}}{C_{ph}},
\end{equation}
where $C_{dc}$ is the number of dark counts in the case of non-illumination, and $C_{tol}$ and $C_{ph}$
are the numbers of total counts and photon counts, respectively, in the case of illumination. $C_{ph}$ is estimated according to
the counts in the main peaks after subtracting the contribution of dark counts, i.e., $C_{dc}$/($f_g$/$f_l$)=$C_{dc}$/100.
$P_{ap}$ as a function of $\eta$ is presented in the inset of Fig. \ref{fig5}(b). For $\eta$=10 \%,
$P_{ap}$ is $\sim$11.7 \%. This moderate afterpulsing probability is expected if one considers the following facts.
Empirically, the SPAD temperature of -50 $^{\circ}$C we used here is somewhat low for rapid gating.
In addition, $P_{ap}$ highly depends on count rate or average time
interval of avalanche detections \cite{GAP09}. Therefore,
considering the parameters of $\eta$ and $\mu$ in the experiment,
11.7 \% is the integration of $P_{ap}$ over $\sim$ 8 $\mu$s,. If we roughly normalize this value to the number of gates during
that average time interval \cite{GAP09}, the afterpulse probability per gate is 11.7 \%/(8 $\mu$s/800 ps) $\sim$ 1.2$\times 10^{-5}$,
corresponding to $6.4\times 10^{-5}$ per ns, which is about 3$P_{dc}$ and consistent with the measurements
shown in Fig. \ref{fig5}. In practice, $P_{ap}$ could be made negligible
by means of applying long logic deadtime \cite{GAP10}.

\begin{figure}[t]
\centering
\includegraphics[width=7.6 cm]{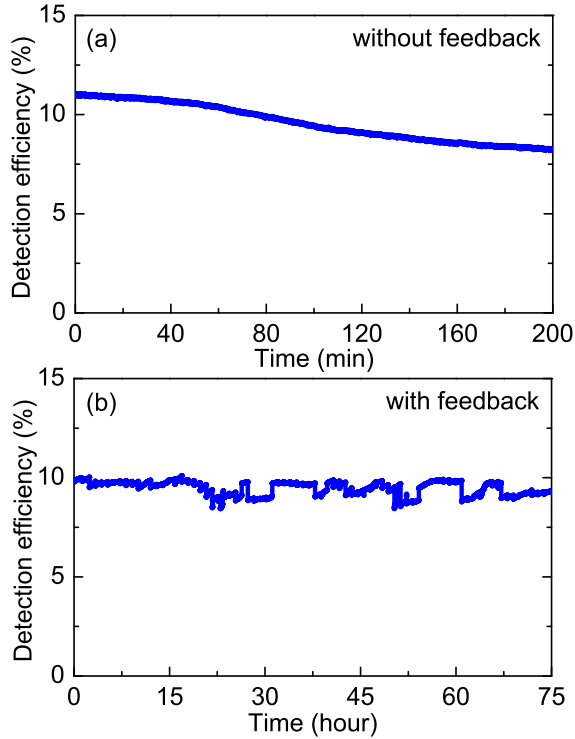}\\
\caption{The system stability test (a) without feedback control for short-term operation and (b) with feedback control for long-term operation.}
\label{fig6}
\end{figure}

Finally, we test the stability performance of the detector system related to
practical applications. If there is no feedback functionality in the detector system, $\eta$
decreases over time due to slight drifts of the relative phase between optical pulses and sine gates.
In Fig. \ref{fig6}(a), $\eta$ is initially 11 \%. It takes 75 minutes for $\eta$ to degrade to
10 \% and 100 minutes to degrade to 9 \%. We designed an automatic peak
search mechanism as feedback control to the detector system. Every 10 minutes of continuous
operation the delay is scanned in the full range and set to the point of maximum count.
This process takes about 1 minute. The detection duty cycle can be improved
by prolonging the feedback time interval and optimizing the feedback algorithm. Using this feedback
control the detector system is continuously operated over 75 hours as shown in
Fig. \ref{fig6}(b). The stability tests show that the detector system is well suited for
practical use, such as in the next generation high-speed QKD applications.
The slight detection efficiency changes in Fig. \ref{fig6}(b) are not due to
the time-division feedback control, but probably to the amplitude fluctuations of sine gates.
In the future, we will apply the technique of automatic gain control for the PA to stabilize the gate amplitude.

\section{Conclusion}

In summary, we have developed a 1.25 GHz gating tabletop InGaAs/InP single-photon detector system that
integrates diverse functions required for practical applications. We have also characterized
the SPAD performance and the long-term stability. Applying such detectors to QKD applications
will significantly enhance QKD performance \cite{GAP10}.

\begin{acknowledgments}
We acknowledge financial support from the CAS, the National High-Tech R$\&$D Program, the National
NSF of China, and the National Basic Research Program.
\end{acknowledgments}

\end{document}